\documentclass[10pt,a4paper]{article}
\usepackage[utf8]{inputenc}
\usepackage{graphicx}
\usepackage{amssymb}
\usepackage{amsmath}
\usepackage{rotating}
\usepackage{datetime}
\usepackage{hyperref}
\usepackage{ulem}
\usepackage{color}

\begin{document}

\begin{center}
\begin{figure}
\hspace{0.5cm} \includegraphics[width=0.1\textwidth]{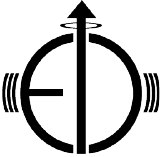}
\end{figure} 
\hspace{2.5cm} \begin{minipage}{14.5cm} 
\vspace{-3cm} 
\begin{center} 
{\Huge \textbf{The Electron-Ion Collider}} \\[0.2cm]
{\LARGE A U.S. facility for the European community \\[0.2cm] 
to explore the mysteries of the building blocks of matter}
\end{center}
\end{minipage}
\rule[0.5cm]{17cm}{0.2pt}



{\Large Contact persons: M. Radici$^1$, S. Dalla Torre$^2$, D. Sokhan$^3$} \\[0.3cm]
{\Large On behalf of the Electron-Ion Collider (EIC) User Group}

\thispagestyle{empty}


\vspace{1cm}
{\bf \large Abstract}

\end{center}

\noindent
{\normalsize This document is submitted as input to the NuPECC Long Range Plan 2024 by three European members of the EIC Users Group Steering Committee (Vice Chair, one “at-large” member, and the EU Representative). We submit the document on behalf of the international EIC Users Group (EICUG) community, but we specifically represent 335 European members of the EICUG (25\%) based in 80 institutions (30\% of the total) located in Armenia, Czech Republic, Finland, France, Germany, Hungary, Ireland, Israel, Italy, Netherlands, Norway, Poland, Slovenia, Spain, Sweden, Switzerland, Ukraine, and the United Kingdom. This European involvement is an important driver of the EIC, but can also be beneficial for a number of related ongoing and planned nuclear physics experiments in Europe. In this document, the shared interest regarding scientific questions and detector R\&D between the EIC and European nuclear physics communities is outlined. The aim is to highlight how these synergies offer ample opportunities to foster progress at the forefront of nuclear physics.}

\vspace{1.5cm}

\begin{figure}[h]
\begin{center}
\includegraphics[width=0.48\textwidth]{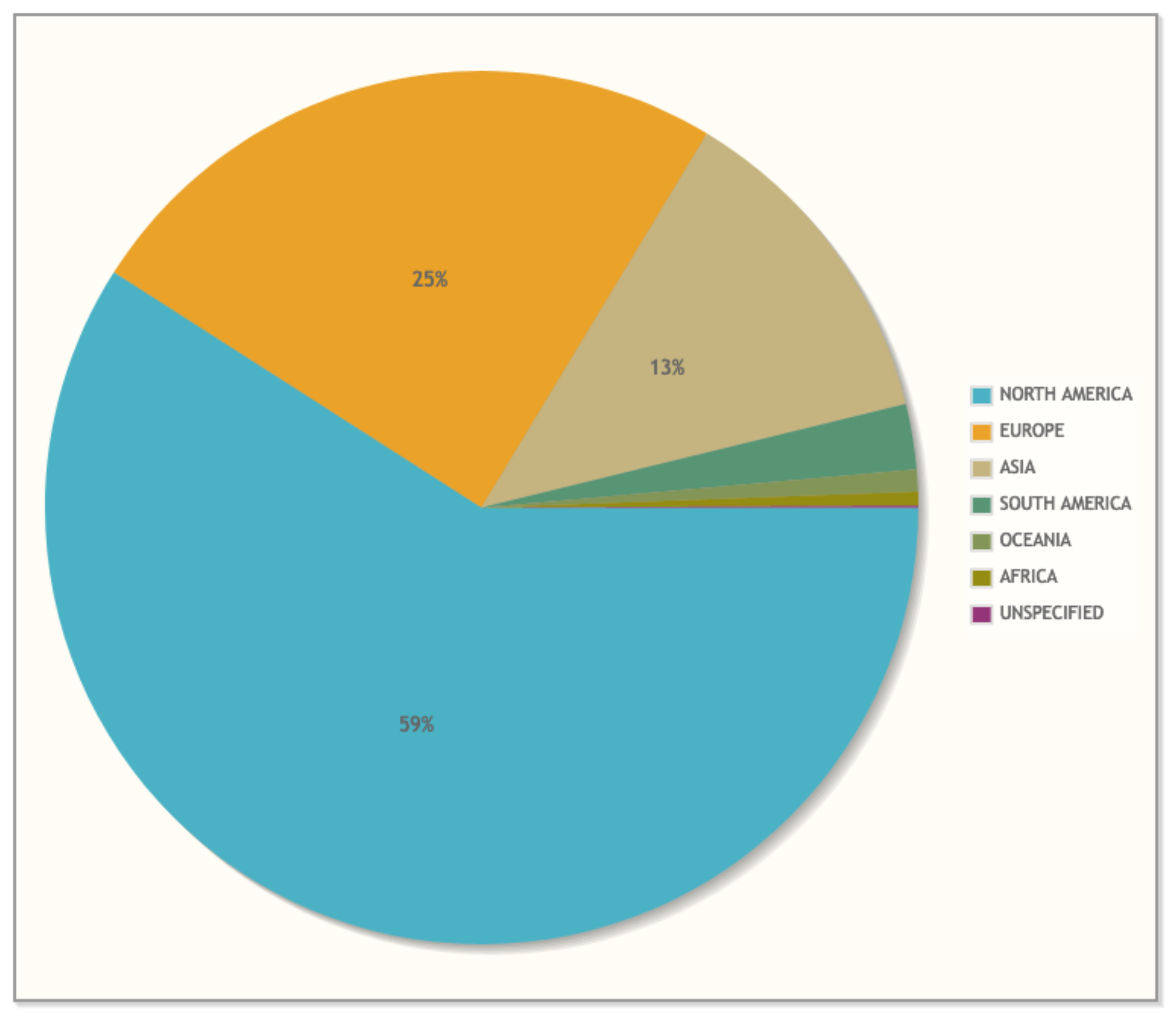} \hspace{0.1cm} \includegraphics[width=0.48\textwidth]{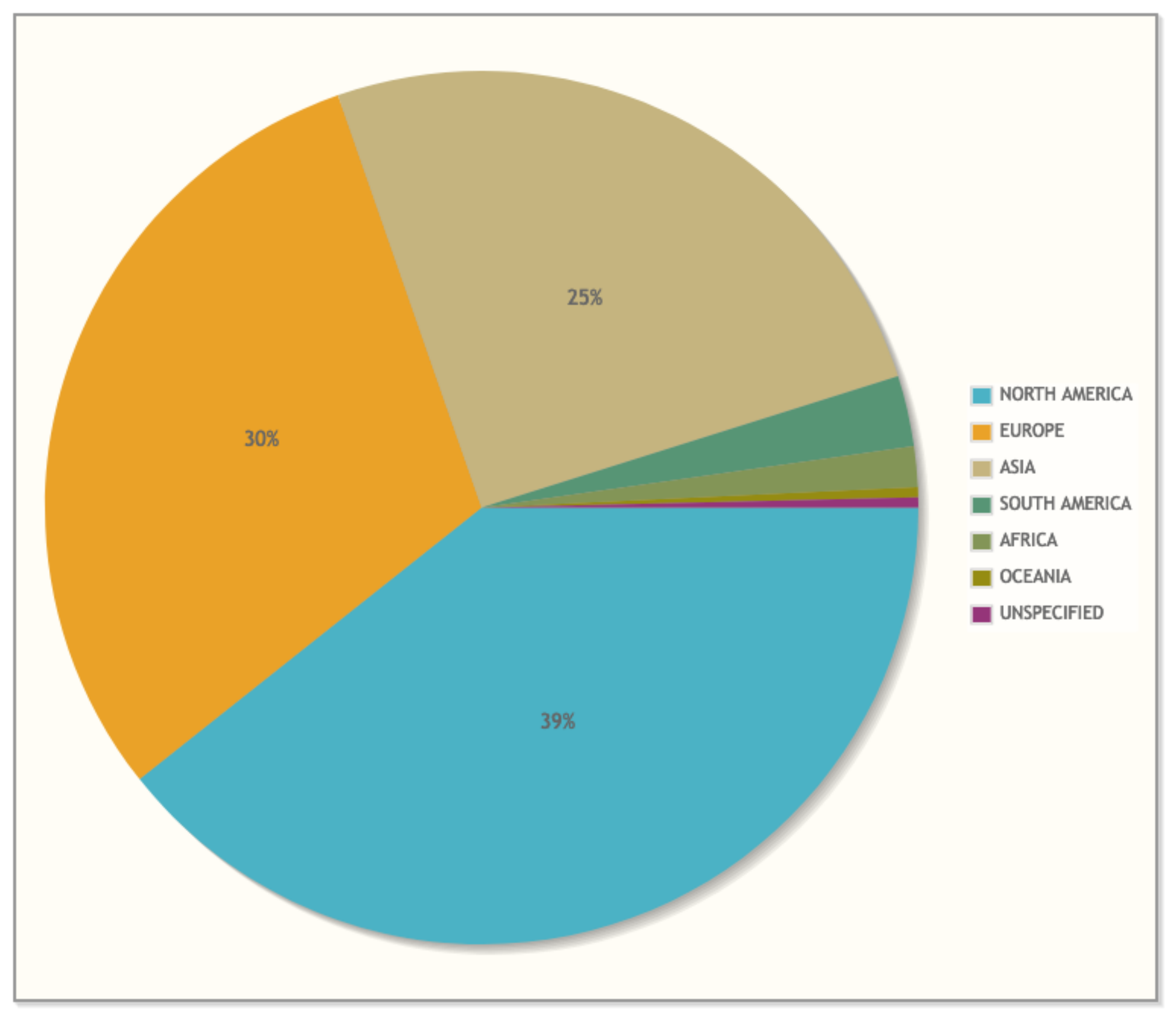}
\end{center}
\end{figure}

\begin{center}
{\normalsize European fraction (orange) of EICUG members (left, 25\%) and institutions (right, 30\%).}
\end{center}

\vspace{1.5cm}
\noindent
\begin{itemize}
\item[$^1$:] {\small INFN Sezione di Pavia, I-27100 Pavia, Italy;  email:} {\tt marco.radici@pv.infn.it}

\item[$^2$:] {\small INFN Sezione di Trieste, I-34149 Trieste, Italy;  email:} {\tt silvia.dallatorre@ts.infn.it}

\item[$^3$:] {\small IRFU, CEA, Universit\'e Paris-Saclay, F91191 Gif-sur-Yvette, France (on leave from University of Glasgow, UK);} \\  {\small email:} {\tt daria.sokhan@cea.fr}
\end{itemize}

\clearpage
\pagenumbering{arabic}

\subsection*{Introduction}
\label{sec:intro}

The Electron-Ion Collider (EIC) is a powerful and versatile new accelerator facility that will be built in the U.S. and is on track to see first collisions in early 2030's. It is capable of colliding highly polarized electrons with high-energy beams ranging from heavy ions to polarized light ions and protons, at a variable center-of-mass energy between 20 and 140 GeV and at a peak luminosity around $10^{34}$ cm$^{-2}$ sec$^{-1}$. 
The EIC will allow for the exploration of new landscapes in QCD, permitting the “tomography”, or high-resolution multidimensional mapping of the quark and gluon components inside of nucleons and nuclei. Because of its unique features, the EIC is designed to be a discovery machine that addresses fundamental questions about how the strong force holds matter together~\cite{WP}. They can be summarized as follows: 
\begin{itemize}
\item[-] How do the properties of nucleons such as mass and spin emerge from quarks and gluons and their underlying interactions?
\item[-] How are quarks and gluons inside nucleons and atomic nuclei distributed in both momentum and position space?
\item[-] How do color-charged quarks and gluons, and jets, interact with a nuclear medium? How do the confined hadronic states emerge from these quarks and gluons? How do the quark-gluon interactions create nuclear binding?
\item[-] What happens to the gluon density in nuclei? Does it saturate at high energy, giving rise to gluonic matter or a gluonic phase with universal properties?
\end{itemize}

To answer these fundamental questions one needs to probe, with high precision, the quark and gluon structure of nucleons and nuclei. It is of great advantage to do this with a simple and well-known probe, such as the electron or the photon. In addition, the high luminosity and cleaner environment (with respect to hadron colliders) will enable precision studies in electroweak physics and some specific searches of physics Beyond the Standard Model (BSM), as well as establish connections to nuclear astrophysics and astroparticle physics. 


The EIC project was recommended in the U.S. 2015 Long-Range Plan for Nuclear Physics {\it ``as the highest priority for new facility construction following the completion of FRIB''}~\cite{NSACLRP}. In 2018, it was endorsed by the U.S. National Academies of Sciences, Engineering, and Medicine (NAS)~\cite{NAS}. In 2019, it was awarded the first project milestone, CD0 or “mission need”, by the U.S. Department Of Energy (DOE), and Brookhaven National Laboratory (BNL) was identified as the construction site. In June 2021, the EIC achieved CD1 status from DOE, marking the start of the project execution phase. In the meantime, the international community released the EIC Yellow Report~\cite{YR}, which condenses the science case and the requirements for a conceptual detector. During the 2022 Summer meeting, the community formed the ePIC (electron-Proton/Ion Collisions) Collaboration around the project of a new general-purpose large-acceptance detector. 
The EIC project is firmly on track towards the start of construction (CD3) in 2025 and the start of operations (CD4) in the early 2030's.

The EIC Users Group (EICUG) is a community of international scientists, currently comprised of approximately 1350 members, a quarter of which are based in Europe (see figure in front matter). 
The regular business of the EICUG is carried out by the Steering Committee, which has a specific European representative and, currently, has two more elected European members: the Vice-Chair and one of the three “at-large” members. In the newly formed ePIC Collaboration, one member of the interim Steering Committee is European, as well as 20\% of the Conveners of all working groups. Thus, the EIC is a U.S. Nuclear Physics project where Europe is playing a key role; it is clearly a project of interest for NuPECC. Moreover, the last European Strategy for Particle Physics (ESPPU) 
final deliberation document~\cite{ESPPU_Final_delib} states that {\it “Electron-proton colliders, such as LHeC or FCC-ep, with the option of including ion targets, are also of interest to NuPECC, which is preparing a support statement for the participation of Europe in the Electron-Ion Collider in the United States”}. Since then, NuPECC installed an EIC task force who initiated an Expression of Interest in the context of Joint ECFA-NuPECC-APPEC Activities~\cite{JENAAEoI}, with the aim of bringing together all interested European physicists. The EIC has also featured prominently in the Horizon2020 STRONG-2020 grant~\cite{Strong2020}, in which a work package was dedicated to detector R\&D for it.

In the following, we outline some of the key topics of the EIC physics case stressing the benefits for the European research in nuclear physics and in neighbouring fields, with which the EIC has many possible synergies~\cite{EICUG2ESPPU}. 




\subsection*{The partonic structure of the Nucleon}
\label{sec:nucleon}

Parton distributions and their multi-dimensional extensions are key to our understanding of the nucleon. They inform about its composition, about its spin (which is still only partially known), about the dynamic generation of its mass from the bare masses of quarks and massless gluons, and the distribution of pressure within it. The latter has implications for our understanding of the structure of neutron stars. While simulation of QCD, the theory of Strong Interactions, on a finite Euclidean discrete space-time lattice is extremely useful, it cannot expose all such details. It is crucial to map out the parton distribution for several key observables: momentum, position and spin, as well as the correlations between them.

The amount of information contained in these maps (or, equivalently, the dimensionality of these maps) depends on the degree of ``exclusivity" of the considered process. In Deep-Inelastic Scattering (DIS) of electrons, one can access the so-called Parton Distribution Functions (PDFs) which are 1-dim objects depending on the fraction $x$ of the nucleon momentum carried by each parton when moving collinearly with the parent nucleon. In Semi-Inclusive DIS (SIDIS), one can also obtain information on the transverse momentum of partons encoded in 3-dim objects called Transverse Momentum Distributions (TMDs). In fully exclusive processes like Deeply-Virtual Compton Scattering (DVCS) or Meson Production (DVMP), one can extract quantities related to the Generalized Parton Distributions (GPDs) which localize partons in the transverse plane at a given $x$. At the EIC, all of these classes of processes can be analyzed with high luminosity, varying the center-of-mass collision energy and making use of the discriminating power that comes with the polarization of both the electron probe and proton / light-ion beams~\cite{YR}. The programme envisaged for the EIC, spanning deep into the gluon sea, is in direct complement to the experiments being currently conducted at BNL and, with strong European participation, at Jefferson Lab and at CERN, and it builds upon the gluon-sea explorations previously conducted at HERA, albeit at much lower luminosity than the EIC. 

Unpolarized PDFs encode key information on the partonic composition of nucleon momentum.  Our knowledge of them is continuously improved by global analyses involving larger and larger sets of data from fixed-target and collider experiments. For example, it has been recently established without statistical ambiguity that there is a charm component in the intrinsic wave function of the proton~\cite{Ball:2022qks}. 
However, PDFs currently give one of the major contributions to the error in precision calculations of Standard Model (SM) observables and in explorations of Beyond SM (BSM) effects. High statistics data obtained with the EIC from neutral and charge currents in electroweak DIS at the highest energy configuration will be able to constrain unpolarized PDFs in regions of low $x$ but also at large $x$, particularly on the largely unconstrained (anti-)strange PDF~\cite{Aschenauer:2019kzf}.

One of the strengths of the EIC is the availability of polarized electron and proton beams, because it allows to do a similar analysis for the polarized quark and gluon PDFs, and thereby to shed further light on their contribution to the proton spin. 
Recent results obtained at RHIC (BNL) give evidence that the gluon contribution is non-vanishing and positive, although the error band is large because the result is very sensitive to the minimum attainable $x$. With its unique capability of 
spanning very small $x$ in polarized collisions, 
the EIC will drastically reduce this uncertainty~\cite{Borsa:2020lsz}. 

When an additional hadron is identified in the final state (SIDIS), the contribution from different quark and antiquark flavors can be separately extracted over a very broad range in $x$. Such analyses go hand in hand with the extraction of new maps, the fragmentation functions (FFs), which describe how a colorless hadron emerges from a colored parton. 
Therefore, our knowledge of PDFs is influenced also by the accuracy at which FFs are known. The FFs are usually extracted from electron-positron collisions, but this yields only a limited knowledge of each individual flavor contribution, and the gluon is reachable only at subleading order. 
At the EIC, the high luminosity, combined with the large lever arm in the hard scale $Q$ of the process and the purposefully planned detector capabilities, will allow for significant improvements for both FFs and PDFs over a wide range of low to medium $x$, including the possibility of studying 
the dependence of the hadronization process on polarization degrees of freedom. This could be accomplished by considering, for example, the fragmentation into polarized $\Lambda$ hyperons, which can usefully complement studies on the production and decay of hyperon-antihyperon pairs foreseen by the {\tt PANDA} experiment at FAIR. 

In SIDIS, one has access also to TMDs, namely to full 3-dim maps of quarks and gluons in momentum space. If the proton and/or the electron are polarized, many new possibilities arise with respect to the collinear framework, depending on the polarization status of the nucleon and the partons~\cite{Bacchetta:2006tn,Angeles-Martinez:2015sea}. All these quantities contain information on the 3-dim orbital motion of quarks and gluons but also on non-trivial dynamical correlations. 

Many critically important aspects of TMDs can be studied experimentally with the EIC. About TMD factorization, an important objective is to demonstrate to what extent the TMDs are universal (like the collinear PDFs and FFs). 
In fact, for specific quark TMDs there is a calculable process dependence~\cite{Collins:2002kn} which is being tested by current experimental facilities with still large uncertainties. The EIC can drastically reduce these uncertainties, and it naturally complements studies of the Drell-Yan process with antiproton beams at lower center-of-mass energies that will be possible at {\tt PANDA}~\cite{Bianconi:2004wu}. For gluon TMDs, which are currently very poorly known, even the unpolarized case is expected to be non-universal. Hence, it is important to compare observables
at hadronic colliders like the LHC to related observables at the EIC~\cite{Boer:2012bt}. 
This comparison will offer a unique opportunity to reach a consistent description of the distribution of partonic transverse momenta over a wide range. While very accurate extractions of unpolarized TMDs~\cite{Bacchetta:2019sam,Scimemi:2019cmh,Bacchetta:2022awv} are available, the TMD framework applies only to low intrinsic transverse momenta: the matching to perturbative QCD calculations at larger momenta is still missing. 
As already outlined, SIDIS at the EIC will allow also to 
explore the flavor dependence of parton intrinsic transverse momenta, with important consequences for the extraction of the $W$ mass from the spectrum of transverse momenta of its decay products~\cite{Bacchetta:2018lna}. 


In exclusive and diffractive processes, one has access to GPDs which encode information about the spatial distribution of partons inside hadrons~\cite{Diehl:2015uka}. This information is independent from the one provided by TMDs because the transverse momentum of partons is not Fourier conjugated to their spatial position. Depending on longitudinal momentum $x$ and transverse position, GPDs give access to the orbital angular momentum carried by quarks and gluons. The detailed measurements that will become possible at the EIC will bring us much closer to a quantitative understanding of the contribution of the partonic orbital angular momentum to the nucleon spin. 

Moreover, GPDs can be connected to linear combinations of the so-called Gravitational Form Factors (GFFs) that parameterize the hadronic matrix elements of the QCD energy-momentum tensor. Through GFF charges, the GPDs will enable studies of the mechanical properties of the nucleon, such as the internal spatial distribution of its charge, energy, pressure, and mass. While the quark contribution to the nucleon mass can be deduced from the $\sigma$ term in $\pi$-N scattering, under some reasonable approximations~\cite{Kharzeev:1995ij} the unknown gluon component can be explored with high precision at the EIC in photo- and electro-production of J/$\psi$ and $\Upsilon$ at threshold~\cite{YR} (see also Ref.~\cite{Lorce:2021xku} for a detailed discussion on the decomposition of the nucleon mass in terms of quark and gluon contributions).  
In an indirect manner, information on spatial distributions of single partons provides also a quantitative baseline expectation to assess correlation effects between different partons, which influences the multiparton dynamics of proton--proton and heavy-ion collisions. While our theoretical knowledge of GPDs has reached a high level of sophistication, progress on relevant measurements, particularly in the medium- and small-$x$ regions, has been comparatively slow because of the required high statistics and excellent detector coverage. The capabilities of the EIC will open the way to a thorough exploration of GPD properties.

\subsection*{From Nucleons to Nuclei}
\label{sec:nuclei}

The non-trivial differences in the $x$-profile of the DIS cross section between a nucleus and a free nucleon is a clear evidence that the nucleus cannot be modeled as a simple superposition of quasi-free nucleons, namely that nuclear PDFs are different from free-proton PDFs. These nuclear modifications are commonly described as shadowing (for $x \lesssim 0.1$), anti-shadowing ($0.1 \lesssim x \lesssim 0.3$) and the nuclear EMC effect ($0.3 \lesssim x \lesssim 0.7$)~\cite{Norton:2003cb}. 
Understanding the exact nature of the mechanisms behind these phenomena is a field actively pursued in both theory and experiment. Hints of a strong link between the EMC effect and nucleon modifications induced by short-range correlations of high-momentum nucleon pairs (SRC) have been suggested~\cite{CLAS:2019vsb}, but alternative explanations exist. In any case, the observation of universal SRC-driven nuclear effects across a wide range of nuclei would provide a definitive explanation for the EMC puzzle, as well as  for other open problems like, e.g., exploring the possible relation between SRC and gluon shadowing. 

Being the first ever collider for DIS on a large variety of nuclei, the EIC 
will enable an unprecedented systematic study of the internal structure of heavy ions in terms of elementary quarks and gluons. These capabilities will impact our currently poor knowledge of nuclear PDFs~\cite{Aschenauer:2017oxs}, and in turn also the heavy-ion studies at hadron colliders that largely depend on the uncertainty of nuclear PDFs, particularly at small $x$. The most prominent example is the challenging problem of understanding how the gluons and quarks from the colliding nuclei form a thermalized plasma. A correct interpretation of observed phenomena in heavy-ion collisions, such as elliptic flow and the ridge, requires an accurate description of the earliest stages of the collision, the initial conditions from which the system evolves into 
the quark-gluon plasma. Here, the EIC will play a crucial role because it will provide the initial picture of the small-$x$ degrees of freedom that constitutes the necessary baseline to study these collective behaviors. Similarly, exclusive photo-production of $J/\psi$ and $\Upsilon$ will give access to gluon PDFs at small $x$~\cite{ALICE:2013wjo}. The possibility to separately perform coherent and incoherent measurements will give an additional handle on probing the nuclear geometry and its fluctuations, complementing similar studies by {\tt LHCb, CMS} and {\tt ALICE} in ultra-peripheral heavy-ion collisions at the LHC. Moreover, accurate measurements of charm structure functions at the EIC would also improve the determination of the nuclear gluon PDF at large $x$, shedding light on the nuclear (anti-)shadowing and EMC effects at medium to large $x$~\cite{YR}.

At high collision energies, or equivalently small $x$, the phase space available for emitting soft gluons is very large leading to the growth of the gluon PDF with decreasing $x$. However, unitarity of the cross section is preserved by the onset of saturation in the gluon distribution due to nonlinear interactions producing gluon merging, potentially forming a new state of matter known as the Colour Glass Condensate. Gluon saturation, therefore, controls the physics in DIS collisions in the regime of small $x$ and moderate scales $Q$. The same kinematics describes the gluons responsible for the initial conditions for quark-gluon plasma production in heavy-ion collisions that are studied at the LHC by the {\tt ALICE, ATLAS, CMS} and {\tt LHCb} Collaborations. While 
there are mounting indications of saturation in past and present experiments, 
at the EIC a high atomic number $A$ can increase the gluon density at a given $Q$, enabling gluon saturation to be more accessible in DIS collisions at lower energies with nuclei than with protons. In order to fully understand this regime, it is important to simultaneously measure inclusive and semi-inclusive cross sections, inclusive diffraction and exclusive reactions, such as diffractive vector meson production~\cite{WP} and dijet correlations~\cite{Aschenauer:2017jsk}. The EIC is being designed to be a facility that can perform this broad set of measurements that are complementary to inclusive particle production and two-particle correlations studies at hadron colliders such as RHIC and LHC. 

The capability of the EIC to produce SIDIS events on nuclei will also provide a laboratory to study the propagation through nuclear matter of high-energy partons, with parton radiation patterns (fragmentation) and formation of parton showers into bound state hadrons (hadronization). 
Such measurements are complementary to similar studies at hadron colliders, and represent the necessary calibration and crosscheck of theoretical approaches to jets and other hard probes that are used to explore nuclear matter properties, such as jet energy loss, jet substructure, medium modification of jets, etc.. In particular, detection of inclusive hadron-in-jet events will complement studies on (un)polarized TMD FFs in vacuum. 

Semi-inclusive and exclusive measurements on nuclei at the EIC will also give access to the essentially unknown TMDs and GPDs of nuclei, respectively. In particular, the possibility of having polarized light-ion beams like 
$^3$He, $^3$H and $^7$Li, will be an extremely useful tool to study 
the polarized EMC effect and to measure the neutron structure function in order to isolate different flavors of the nucleon TMD~\cite{YR}.

\subsection*{Connections with other fields}
\label{sec:outside}

The science program of the EIC can have synergies with nuclear and particle astrophysics. The ability to map the pressure distributions in the nucleon, arising from the relationship between GPDs and GFFs, has implications for our understanding of the structure of neutron stars. Nuclear PDF errors currently form the dominant theory error in the description of high-energy neutrino-nucleus cross sections that are needed for studies of cosmic rays~\cite{AbdulKhalek:2022hcn}. Improving our knowledge of nuclear PDFs at small $x$ at the EIC would be very beneficial to such studies.

The EIC can potentially impact also searches of new BSM physics. In several instances, these explorations rely on an accurate determination of PDFs at large $x$. Similarly, it is crucial to know a very precise value of the $W$ mass, which however needs a careful study of the uncertainty induced by the flavor dependence of parton intrinsic transverse momenta encoded in TMDs~\cite{Bacchetta:2018lna}. 
Thanks to the high luminosity, at the EIC it will be possible to reach a precise measurement also of the weak mixing angle $\sin^2 \theta_W$ through collisions of polarized electrons and unpolarized deuterons at an intermediate $Q$ range, which is beyond the reach of fixed-target experiments~\cite{YR,AbdulKhalek:2022hcn}. Deviations from the predicted SM running would signal BSM physics, particularly at low energy through effects induced by ``dark'' $Z$ bosons. 
Finally, the EIC measurements can constrain particular directions in the parameter space explored by SM Effective Field Theories which are not accessible at the LHC. For example, possible new tensor interactions at the TeV scale can be better constrained by extracting the nucleon tensor charge. At the EIC, SIDIS measurements involving transversely polarized protons will contribute to significantly reduce our uncertainty on the extraction of the proton tensor charge by widely enlarging the covered $x$ and $Q$ range~\cite{YR}. 




\subsection*{Detector R \& D}
\label{sec:detector}

The broad EIC physics program encompasses the study of inclusive, semi-inclusive and exclusive reactions in the collision of electrons with protons or light to heavy nuclei over a wide range of center-of-mass energies. 
The challenge, experimentally, is to design a large acceptance, general detector that is capable of reconstructing this wide variety of channels. The detector must provide coverage for a wide range of center-of-mass energies, and must take full advantage of the collider high luminosity, the beam polarization options and the variety of species of the ion beam, opening an era of precision measurements in deep inelastic scattering.

The detector concept has been developed over years in parallel with the definition of the physics case. The progress has been marked by several milestones, where the most recent ones are the EICUG Yellow Report~\cite{YR} and the Detector Proposals in response to a corresponding call by the EIC Project~\cite{DetProp}. 
These efforts have culminated in the design of the project detector ePIC. The community has expressed strong support for a second detector that would be complementary to ePIC and perhaps utilize alternative technologies and interaction region components. 

The key detector requirements and corresponding reference technologies are well identified and briefly summarized here.
The unambiguous identification of the scattered electron and the precise measurement of its angle
and energy is essential, since it determines the key kinematic variables $(x,Q)$ of the interaction. 
This is achieved with precision tracking using a high-resolution electromagnetic calorimeter system and by minimizing the material budget of the tracking system in front of the calorimeter. 
Magnetic analysis is provided by a 1.7~T solenoid, while the tracking system is based on vertexing and tracking layers using low-mass, low-power consumption MAPS in 65~nm technology, complemented by large area MPGDs. Electromagnetic calorimetry with finest resolution is provided in the backward endcap by lead tungstate crystals, while sampling calorimetry with scintillating fibers and tungsten powder is adopted in the forward endcap. Scintillating glass blocks with projective geometry form the barrel calorimeter, where a hybrid calorimeter with imaging layers followed by a sampling section is being considered as a potential alternative.

The semi-inclusive processes require excellent hadron identification over a wide momentum and rapidity range, from 200 MeV/c to 10 GeV/c in the barrel region and up to 50 GeV/c in the forward (hadron going) region, with full 2$\pi$ acceptance for tracking and momentum analysis and excellent vertex resolution by a low-mass vertex detector. In particular, experiments at the EIC specifically require hadron identification over an extremely wide phase-space range and enforce the use, within the same setup, of diversified approaches for particle identification. The PID system includes Cherenkov devices: a dual radiator (aerogel and gas) RICH in the forward endcap, a high performance DIRC in the barrel, and an aerogel RICH in the backward endcap, either with modular units using Fresnel lenses or in a proximity focusing configuration with a long proximity gap. Low-momentum particle PID is provided by a time-of-flight system based on AC-LGAD sensors. 

Exclusive reactions require the capability to accurately track all particles with high spatial and momentum resolution. The identification of many key processes depends on the complete hermiticity of the setup. This is a peculiar feature of experiments at the EIC, resulting in the additional requirement of very forward detectors, such as Roman pots, and large-acceptance zero-degree calorimetry
to effectively detect neutrons from the breakup of nuclei. Integration of “tagging” detectors in the far-backward region for the detection of electrons scattered at the lowest angles enables the exclusive reconstruction of photo-production processes. 

The entire experimental program will need a precise measurement of luminosity; polarized beams will also require highly accurate electron, proton, and light--nucleus polarimetry.

A triggerless data acquisition system, configured in streaming read-out mode, is motivated by the precision physics program. A very high average throughput will be achieved using artificial intelligence algorithms on high performance computing platforms. 

The need for a largely diversified set of detector technologies as well as up-to-date approaches for data acquisition and data reduction emerges from this schematic analysis.  Several technologies still require the completion of the R\&D phase and they are of interest for applications beyond the EIC detectors. Related expertise and experience are spread over many communities and geographical areas. Europe has a long tradition and advanced know--how in detector development, together with a set of future projects that look very naturally synergistic with the EIC in the field of detector technology. In particular, in the nuclear sector we have the following: {\tt ALICE} ITS3 upgrade and the further future ALICE3 (both at the LHC at CERN), the {\tt AMBER} experiment at CERN SPS (continuing the {\tt COMPASS} tradition), and the {\tt PANDA} experiment at GSI FAIR. More synergies can be identified by considering also high energy physics, including all the major experiments at the LHC and their upgrades. 

These opportunities are being recognized by the community and some initiatives to build upon them have already taken place~\cite{JENAAEoI,JENAAKickoff,Strong2020,RDday,ALICE3}. At the same time, there 
is already a significant involvement of European physicists in the U.S. generic EIC detector R\&D program. Groups from Armenia, Czech Republic, France, Italy, Poland, and United Kingdom, already collaborate with U.S. groups in the R\&D projects.

\subsection*{Suggested recommendations}
\label{sec:end}

In summary, we can undoubtedly say that Europe is providing a substantial contribution to the development of the EIC science case and of the related detector technologies, thanks to its recognized leadership in several fields. More importantly, this effort should be regarded as a unique opportunity of expanding the European physics program with a large science return for a modest investment. 
Because of the many synergies outlined above, the outcome 
will turn out to be extremely beneficial for the European research itself, particularly in nuclear physics but also in other neighbouring fields, both at the scientific level and for the detectors that will be designed in future European experiments. 


Therefore, we express the hope that the NuPECC Long Range Plan will 
\begin{itemize}
\item acknowledge the large European involvement in the EIC project (roughly speaking, 17\% of the European Hadron Physics community), with interests on topics that are central to current nuclear physics research

\item encourage and actively support, where possible, the participation of members of the nuclear physics community in the EIC project

\item continue and reinforce the scientific and technology activities with strong synergies between the EIC and European nuclear physics projects.

\end{itemize} 

\subsection*{Acknowledgments}
\label{sec:thanks}

We thank Elke Aschenauer, Dani\"el Boer, Rolf Ent, Renee Fatemi, Yuri Kovchegov, Eugenio Nappi, Mariusz Przybycien and Franck Sabati\'e  for revising the draft and making insightful comments and suggestions.

\end{document}